\documentclass{article}

\usepackage{arxiv}

\usepackage[utf8]{inputenc} 
\usepackage[T1]{fontenc}    
\usepackage{hyperref}       
\usepackage{url}            
\usepackage{booktabs}       
\usepackage{amsfonts}       
\usepackage{nicefrac}       
\usepackage{microtype}      
\usepackage{cleveref}       
\usepackage{lipsum}         
\usepackage{graphicx}
\usepackage{doi}
\usepackage{xcolor}

\title{A Data Literacy Competence Model for Higher Education and Research}

\date{\today}

\newif\ifuniqueAffiliation
\uniqueAffiliationtrue

\ifuniqueAffiliation 
\author{ \href{https://orcid.org/0000-0003-3673-3663}{\includegraphics[scale=0.06]{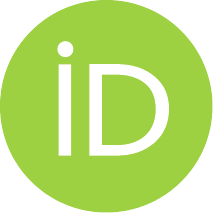}\hspace{1mm}Martina M.~Echtenbruck} \\
	Faculty of Information and communication sciences\\
    Institute for Information Science\\
	TH Köln, Germany \\
	\texttt{martina.echtenbruck@th-koeln.de} \\
 \And
	\href{https://orcid.org/0000-0001-5985-739X}{\includegraphics[scale=0.06]{orcid.pdf}\hspace{1mm}Simone Fühles-Ubach} \\
	Faculty of Information and communication sciences\\
	Institute for Information Science\\
	TH Köln, Germany \\
	\texttt{simone.fuehles-ubach@th-koeln.de} \\
 \And
	\href{https://orcid.org/0000-0002-8969-4795}{\includegraphics[scale=0.06]{orcid.pdf}\hspace{1mm}Boris Naujoks} \\
	Faculty of Computer Science and Engineering\\
	Institute for Data Science, Engineering, and Analytics\\
	TH Köln, Germany \\
	\texttt{boris.naujoks@th-koeln.de} \\
 \And
	\href{https://orcid.org/0009-0001-0495-5125}{\includegraphics[scale=0.06]{orcid.pdf}\hspace{1mm}Elisabeth Kaliva} \\
	Faculty of Cultural sciences\\
	Köln International School of Design (KISD)\\
	TH Köln, Germany \\
	\texttt{elisabeth.kaliva@th-koeln.de} \\
}
\else
\usepackage{authblk}

\newbox{\orcid}\sbox{\orcid}{\includegraphics[scale=0.06]{orcid.pdf}} 
\author[1]{%
	\href{https://orcid.org/0000-0000-0000-0000}{\usebox{\orcid}\hspace{1mm}David S.~Hippocampus\thanks{\texttt{hippo@cs.cranberry-lemon.edu}}}%
}
\author[1,2]{%
	\href{https://orcid.org/0000-0000-0000-0000}{\usebox{\orcid}\hspace{1mm}Elias D.~Striatum\thanks{\texttt{stariate@ee.mount-sheikh.edu}}}%
}
\affil[1]{Department of Computer Science, Cranberry-Lemon University, Pittsburgh, PA 15213}
\affil[2]{Department of Electrical Engineering, Mount-Sheikh University, Santa Narimana, Levand}
\fi

\renewcommand{\headeright}{Technical Report}
\renewcommand{\undertitle}{Technical Report}
\renewcommand{\shorttitle}{A Data Literacy Competence Model for Higher Education and Research}

\hypersetup{
pdftitle={A template for the arxiv style},
pdfsubject={q-bio.NC, q-bio.QM},
pdfauthor={David S.~Hippocampus, Elias D.~Striatum},
pdfkeywords={First keyword, Second keyword, More},
}

\begin{document}
\maketitle

\begin{abstract}
    In an increasingly data-driven world, the ability to understand, interpret, and use data - data literacy - is emerging as a critical competence across all academic disciplines. The Data Literacy Initiative (DaLI) at TH Köln addresses this need by developing a comprehensive competence model for promoting data literacy in higher education. Based on interdisciplinary collaboration and empirical research, the DaLI model defines seven overarching competence areas: "Establish Data Culture", "Provide Data", "Manage Data", "Analyze Data", "Evaluate Data", "Interpret Data", and "Publish Data". Each area is further detailed by specific competence dimensions and progression levels, providing a structured framework for curriculum design, teaching, and assessment. Intended for use across disciplines, the model supports the strategic integration of data literacy into university programs. By providing a common language and orientation for educators and institutions, the DaLI model contributes to the broader goal of preparing students to navigate and shape a data-informed society.
\end{abstract}


\section{Introduction}
As our world becomes increasingly digital, more and more processes and activities, essential aspects of everyday life, research, and work, are taking place digitally. All kinds of data are constantly analyzed and processed to enhance our everyday experiences. One example is health data, information like sleep patterns, exercise routines, and food intake, can be continuously tracked via smartphones and smartwatches. Utilizing data analytics, insights are used to refine training schedules and offer personalized nutrition plans.
As a result of this digitalization, the amount of available data is rapidly growing. Research is becoming increasingly data-intensive \cite{Hey2006, Lynch2009}. This opens up great potential but also brings the need to face entirely new challenges.

Interdisciplinary knowledge of how to deal with data in a planned and secure manner and its consciously and ethically appropriate use is becoming crucial for almost all areas of business, as well as the everyday social live of all people. Therefore, data literacy, i.e., the ability to collect, visualize, critically evaluate, and consciously use data will become a core competence across all disciplines (cf. \cite{Ridsdale2015}). It is essential to teach and sharpen comprehensive knowledge and skills in data literacy at all levels of education.

Data literacy encompasses the competencies required for conscious and responsible data handling. However, there are various definitions of data literacy in the literature. Swan et al. e.g., state that ``data literacy is the ability to ask and answer meaningful questions by collecting, analyzing and making sense of data encountered in our everyday lives'' ~\cite{Swan2009}.
Johnson defined data literacy as the ability to process, sort, and filter vast quantities of information, which requires knowing how to search, filter and process, and produce and synthesize it \cite{Johnson2012TheID}. This definition was later adapted from Koltay \cite{Koltay2016}. 
Shortly afterward, ACRL \cite{ACRL2013} stated that ``students, both as users and as future creators of data, should be trained to understand how their choices affect access, reuse, and preservation; libraries are better placed than any other academic unit to carry out that training''.
Eventually, Ridsdale et al.~\cite{Ridsdale2015} put it as follows: “Data literacy is the ability to deal with data in a planned way and to be able to collect, manage, evaluate and apply it in a critical way in the respective context”. 

The TH Köln's Data Literacy Initiative (DaLI) was established to create a modular, interdisciplinary certificate program open to all students of TH Köln. The initiative seeks to systematically embed data literacy across all facets of teaching and research within the university. However, this diversity of definitions creates a challenge in clearly conceptualizing and universally applying the concept of data literacy. A guiding definition of the term 'data literacy' and a well-defined competence model or framework resp. are necessary to classify skills, recognize and credit them within an education or certificate system, and properly teach what is understood by data literacy. 
Hence, this article aims to specify a competence model that enables teaching ethically and legally sound data competencies.

In the following sections, we will first explore the concepts of data, competence, and the data lifecycle (Sec.~\ref{sec:preliminaries}). Subsequently, we will examine different definitions of, as well as competence models for data literacy (Sec.~\ref{sec:previousWork}). Following this, drawing on these insights, we will formulate our data literacy competence model in Section~\ref{sec:DaLI-Model}. 
Finally, in Section~\ref{sec:conclusion}, results are discussed.

\section{Preliminaries}\label{sec:preliminaries}
The following serves as an introductory exploration of basic concepts and terminology essential for illustrating the proposed competence model. We begin by exploring the concept of data and then consider different (research) data life cycles. Finally, we examine the notions of literacy and competence in order to provide a comprehensive basis for our discussion.

We come into contact with data incessantly, and the word data is regularly used in everyday life. But what is data? The Cambridge Dictionary states that data is: “information, especially facts or numbers, collected to be examined and considered and used to help decision-making or information in an electronic form that can be stored and used by a computer; information, especially facts or numbers, collected to be examined and considered and used to help with making decisions; information in an electronic form that can be stored and processed by a computer”~\cite{CambridgeDictionary}.
Although this is already a fairly broad definition, the Merriam-Webster Dictionary adds: “[...] information in digital form that can be transmitted or processed; information output by a sensing device or organ that includes both useful and irrelevant or redundant information and must be processed to be meaningful”~\cite{MerriamWebster}.

This does not provide an easy-to-grasp definition of data. However, it should be noted that data is a multifaceted term that needs to be reflected accordingly in the definition of competencies.
What can already be read here, however, is that data can be collected, processed, used for decision-making, and stored. Data, therefore, follows a specific data life cycle. A well-defined data lifecycle provides a sound basis for developing a competence model, as it already provides a specific structure and allows the individual competencies to be aligned with it.

Many similar definitions exist for the lifecycle of data~\cite{NFDI4Chem,HKLib, FDI}. Only a few stand out due to their special features. J.M. Wing~\cite{Wing2018}, for example, presents a very detailed data life cycle. She proposes considering the following stages: ‘Generation’, ‘Collection’, ‘Processing’, ‘Storage’, ‘Management’, ‘Analysis’, ‘Visualization’, ‘Interpretation’ and even ‘Human’. 

Some parts of data processing are particularly important in higher education and research. Data generated or used as part of the research process is referred to as research data. It comprises factual material generally recognized in the scientific community as necessary for validating research results~\cite{Circ110}. Research data often represents reality at a specific point in time and is, therefore, often singular and difficult or costly to recover~\cite{MinWFK2014}. In addition, the reproducibility of research and its results is crucial for its integrity and trustworthiness. Therefore, appropriate archiving, publication, and the possibility of reusing data are especially important in research.
As a result, many higher education institutions worldwide have sought to develop a dedicated research data lifecycle that integrates the specific concerns of academics and their research data~\cite{UWM2022,HMS,UBDKF2022,CTU2024,FUB}.

The RWTH Aachen~\cite{RWTH2021} proposes a research data lifecycle that comprises 6 stages: Planning, collection, analysis, archiving, access, and re-use. The work of the TH Köln DaLi Project is based on this last data life cycle. However, we replaced the last three stages used by RWTH Aachen with the two stages: "Data Sharing and Publishing" and "Data Preservation and Reuse" (cf. Figure 1). Sharing and publishing of data and results is usually done once towards the end of the project. However, long-term archiving and maintenance of data to enable re-use is an ongoing process that continues after publication and project completion. Additionally, the actual reuse of the data starts a new lifecycleof its own. 

\begin{figure}[!htb]
  \centerline{\includegraphics[width=0.4\textwidth]{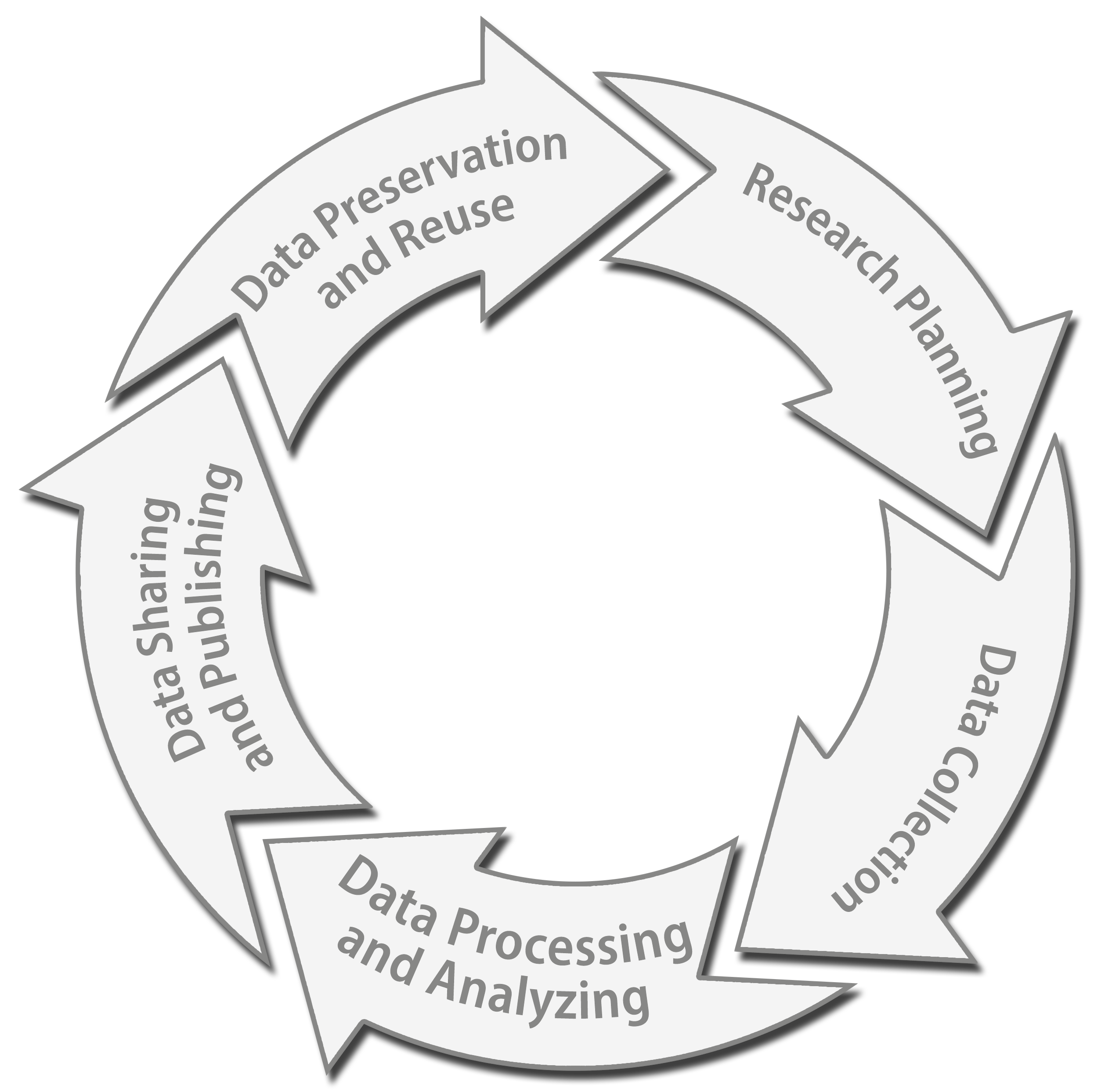}}
  \caption{The (research-) data lifecycleas referred to by DaLI (TH Köln)}
  \label{fig:datalife cycle}
\end{figure}

With the data lifecycle providing the necessary structure for the competence model, the question of how best to specify the required competencies remains. So what exactly are competencies?

Weinert~\cite{weinert2002} explains the concept of competencies as "the cognitive abilities and skills available in individuals or learnable through them to solve certain problems, as well as the associated motivational, volitional and social willingness and ability to solve problems successfully and responsibly in variable situations". According to this definition, competence includes knowledge, the ability to apply this knowledge (skills), and the willingness to do so (attitude). 

Schott and Ghanbari consider it problematic to include values such as willingness in the definition of competence as it unnecessarily restricts the scope of competence. You can have or acquire a skill that you detest but are still good at, and it can even be useful, e.g., doing a tax return~\cite[p. 26f]{Schott2012}. Instead, they state that competence is an ability characterized by a certain sustainability and persists as a characteristic of a person over a longer period. It is described by two statements~\cite[p. 38]{Schott2012}:
\begin{itemize}
    \item [1.] indication of a specific set of tasks that one can perform if one possesses the competence in question; this set of tasks may include subsets of different types of tasks and 
    \item [2.] indication of a level of competence, or, in the case of several subsets of tasks, several levels of competence, which determine how well one can perform the tasks in question if one possesses the competence in question".
\end{itemize}

They state that this concept of 'competence' is broader than the concepts of competence currently used in education. "Our concept of competence encompasses any type of ability that is the subject of teaching and learning in education, regardless of the type of ability in question. Accordingly, competencies can be, for example, factual knowledge, problem-solving, creativity, cognitive, social, emotional and motor skills"~\cite[p. 38]{Schott2012}.

However, we also believe raising awareness and reflecting on attitudes is important, especially regarding ethical and privacy issues. We thus include awareness about these issues as knowledge items in the proposed DaLI Competence Model.
Therefore, to formulate our competence model, we focus on defining knowledge goals and tasks that can be mastered with the corresponding competence. However, a further breakdown of the named tasks for mapping competence levels is not provided below.

\section{Previous and Related Work}\label{sec:previousWork}
Data literacy encompasses a wide range of competencies and skills. While various definitions of data literacy exist, only a few competency models or frameworks have been formulated, given that data literacy in education is still a relatively new development. In the following, we will take a closer look at some of the existing competence models and frameworks before formulating our own competence model.

The work of Ridsdale et al.~\cite{Ridsdale2015} may well be the best known in this area. Ridsdale et al. define data literacy as “the ability to collect, manage, evaluate, and apply data in a critical manner”. As they state, this definition is based on an extensive analysis of relevant literature sources and existing definitions. They also state that data literacy “is an essential ability required in the global knowledge-based economy; the manipulation of data occurs in daily processes across all sectors and disciplines. An understanding of how decisions are informed by data, and how to collect, manage, evaluate, and apply this data in support of evidence‑based decision‑making, will benefit Canadian citizens, and will increasingly be required in knowledge economy jobs”. 
Based on their definition, they specified five core aspects of data literacy (Conceptual Framework, Data Collection, Data Management, Data Evaluation, and Data Application) and mapped the individual abilities and skills to these core aspects. They identify a total of 64 abilities and skills, which in turn are divided into three different levels (Conceptual Competencies, Core Competencies, and Advanced Competencies).

The Georg-August-University of Göttingen has established a program for teaching data competencies on campus as part of the project “Daten Lesen Lernen”~\cite{UniGoettingen}. In defining their competency framework, they followed the work of Ridsdale et al. and the strategy for “Bildung in der digitalen Welt” (Education in the digital world) formulated by the Standing Conference of the Ministers of Education and Cultural Affairs~\cite{KMK2017,KMK2016}.
They state that “the term data literacy covers a variety of skills, such as the ability to investigate and analyze data in an explorative, computer-aided manner and to interpret the results obtained. Data literacy thus transcends traditional subject boundaries and integrates skills from the fields of computer science, statistics, mathematics with ethical and social science aspects as well as detailed knowledge from the area of application under consideration.”~\cite[par. 2]{UniGoettingen}.
Taking this as their underlying definition, it is somewhat less concrete than the definition of Ridsdale et al.. However, they also consider the ethical and social aspects of data literacy.
For the “Daten Lesen Lernen” project, a competency framework has been specified that aligns the competencies that can be acquired in the area of data literacy via this program along two dimensions. The first dimension matches the core aspects of data literacy as specified by Ridsdale et al., but the ‘Conceptual Framework’ was omitted. The other dimension corresponds with the fields of competence as specified in~\cite{KMK2016}. In the remainder, this project is referred to as Göttingen.

In their work, Heidrich et al.~\cite{heidrich2018} focus on best practice examples for cross-disciplinary knowledge transfer in the section on data literacy. The question "What is data literacy?" is one of the guiding questions of their work. To answer this question they conducted a comprehensive study. 
One of the interesting findings is that most experts generally agree with Ridsdale's definition of data literacy, only additions are requested. The missing topics include 'empathy with data use', 'contextualisation', 'basic understanding', 'visualisation', 'legal aspects', 'modelling', and 'interpretation'. 
The individual topics were also covered with varying frequency in the data literacy training programmes available at the time of Heidrich et al.'s research, with topics such as 'data preservation', 'data healing, security and reuse', and 'data citation' being covered by less than 30\% of the courses.
Overall, the survey results show great uncertainty in the definition of data literacy. However, it also identifies many aspects that should be considered when defining a competence model for data literacy.

A very detailed framework for data literacy has been formulated by Schüller et al.~\cite{Schueller2020}. 
They first reflect on the concepts of ‘competence’ and ‘competence framework’ before they start formulating their competence framework. To define the concept of 'competence', they refer to Weinert ~\cite{weinert2002}. They note that this reference definition includes not only knowledge and the ability to apply this knowledge (skills), but also the willingness to do so (values). They express that these three dimensions of competence should thus also form dimensions of a competence framework.
To specify what is covered by the term data literacy, they refer to Heidrich et al. and Ridsdale et al.. Additionally, they state that “data literacy can be understood as the ability that a responsible citizen needs in today’s society to find his way through an overabundance of data and information and to make informed decisions – in everyday life and at various political levels. This ability to make decisions requires the ability to distinguish data and information from interpretations and opinions.” ~\cite[p. 11]{Schueller2020}.

The Data Literacy Framework of Schüller et al. is structured into four levels~\cite[p. 23ff]{Schueller2020}:

The first level comprises productive and receptive competence fields, including establishing a data culture, providing data, evaluating data, interpreting results, interpreting data, and deriving actions.
The second level provides a more detailed summary of the competences within each competence field, such as identifying data applications, modeling data applications, data analysis, data visualization, and others.
To enable a comprehensive presentation of the competencies, the third level lists the dimensions: knowledge, skills/abilities, as well as motivation and (value) attitudes. The fourth level defines competence levels as "basic level," "advanced level," and "expert level."

The Data Literacy Framework of Schüller et al. encompasses 18 competencies essential for decision-making and knowledge development in data-related contexts. The cyclical process model categorizes the relevant process steps and associated competencies into productive and receptive components. The productive area involves competencies necessary for deriving data products from the available data. On the other hand, the receptive area encompasses competencies and tasks relevant for decoding data projects and uncovering the underlying data. Consequently, data literacy encompasses both the skillful and professional creation of data products and the competent and reflective handling of data, which includes interpretation, description, and application of the data by the users.

In the proposed DaLI Competence Model, we include significant aspects from Göttingen and Schüller et al. that are not covered by the work of Ridsdale et al., and we made adjustments to meet the needs of higher education and the scientific community. However, since Ridsdale's competence model is well-founded and cited in most subsequent work, we used this model as the main basis for our work. In the following we present the DalI Competence Model in detail.
\section{The DaLI Competence Model: Derivation of a structured model for new forms of interdisciplinary knowledge transfer in the field of data literacy}\label{sec:DaLI-Model}
\begin{figure}[!htb]
  \centerline{\includegraphics[width=0.8\textwidth]{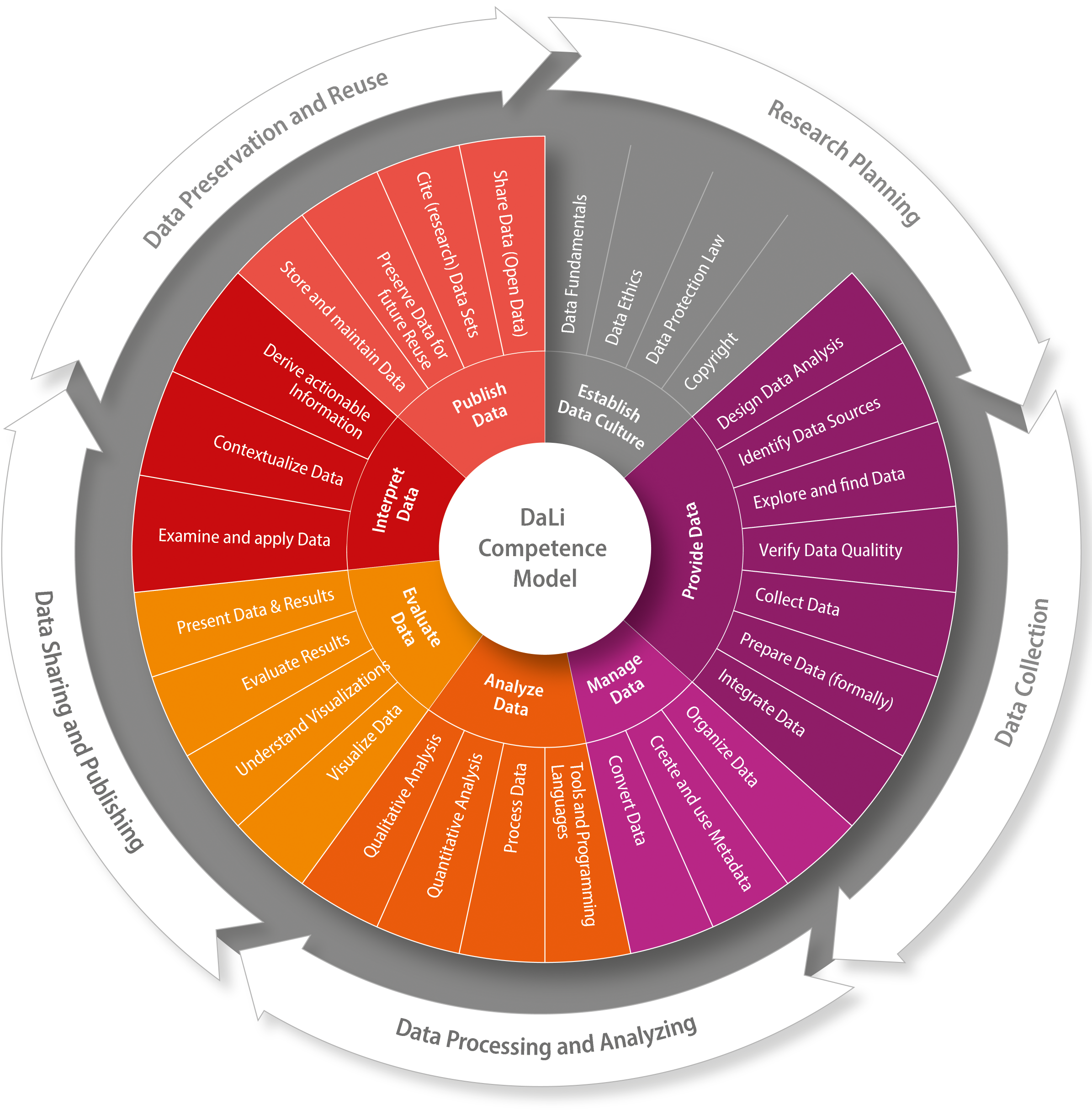}}
  \caption{The proposed DaLI Competence Model with its seven competence areas, the associated competencies, and the corresponding Research Data Life Cycle as proposed by DaLI (TH Köln).}
  \label{fig:competenceModel}
\end{figure}

Data literacy encompasses a wide range of competencies, which in the model presented here, are organized into seven areas, namely: ‘Establish Data Culture’, ‘Provide Data’, ‘Manage Data’, ‘Evaluate Data’, ‘Interpret Data’, ‘Understand Data’, and ‘Publish Data’. In the graphical representation of our competence model, these competence areas are organized in a circle based on the data's life cycle (cf. Figure~\ref{fig:competenceModel}).

A particular focus is on the area of ‘Establish Data Culture’, which also forms the basis of the model. The graphic representation emphasizes this competence area through its coloring, size, and positioning.
This emphasis is intended to indicate that conscious and responsible use of data, particularly critical, ethical, and privacy-compliant use, is to be promoted in all areas of the DaLI Competence Model.

The DaLI Competence Model is an essential tool to foster data literacy. Integrating the research data lifecycle ensures that the acquired competencies are practically applicable in higher education and research. In addition, the emphasis on critical thinking, ethical reflection, and data protection skills in the "Establish Data Culture" area cultivates responsible and adept data users, preparing them for academic and professional endeavors.

The individual areas of competence with examples of associated knowledge items and tasks are presented and discussed in detail below.

%

\subsection{Establish Data Culture}\label{sec:EstablishDataCulture}

The ‘Establish Data Culture’ competence area is the first area in our competence model. With its competencies ‘Data Fundamentals’, ‘Data Ethics’, ‘Data Protection Law’ and ‘Copyright Law’, it not only covers the first topics related to a basic understanding of data, but also forms a foundation of crucial competencies that should be considered when dealing with data.

The competence of ‘Data Fundamentals’ encompasses a foundational understanding of data as a valuable resource that enables informed decision-making. It encourages the sharing of data and insights, thereby facilitating collaboration and promote innovation. Furthermore, the proposed model's ‘Data Fundamentals’ competence incorporates principles of open science~\cite{Heise2020}, encompassing knowledge of the data lifecycle, open access principles~\cite{Suber2012,Swan2012}, open data practices~\cite{MurrayRust2008}, adherence to FAIR principles~\cite{Wilkinson2016,NG2017a}, as well as the ability to apply and critically reflect upon them. We see these elements as core to a data culture we want to foster.

‘Data Ethics’ ensures that decisions are made in accordance with moral principles. Knowledge of Data Ethics is essential for making informed, responsible decisions regarding data use, safeguarding individuals' privacy, promoting fairness and transparency, and ensuring that data practices align with ethical standards and societal values.

The competence ‘Data Protection Law’ protects the privacy and rights of individuals. Knowledge of ‘Data Protection Law’ is essential to ensure compliance with legal requirements, protect individuals' personal information, maintain organizational integrity, and build trust with stakeholders.

‘Copyright’ ensures that individuals have the necessary knowledge to make informed decisions on how to use copyrighted materials. Understanding copyright laws and regulations is essential to avoid infringement and ensure the ethical and legal use of data. Additionally, knowing about licensing helps when it comes to publishing data and research results at the end of a project, ensuring that intellectual property rights are respected and that the dissemination of research is conducted in compliance with legal standards.

Though previous models have also addressed the topics of data culture, data ethics, and data protection law, the proposed model emphasizes these concepts. Copyright is introduced as a new component. These aspects are essential to ensure that data is not only used effectively but also handled ethically, proactively, and in compliance with the law. They are fundamental prerequisites for a conscious and responsible use of data.
The competencies assigned here affect and influence all other competence areas and are considered and repeatedly addressed in the following competence areas. 

\begin{figure}[!htb]
  \centerline{\includegraphics[width=\textwidth]{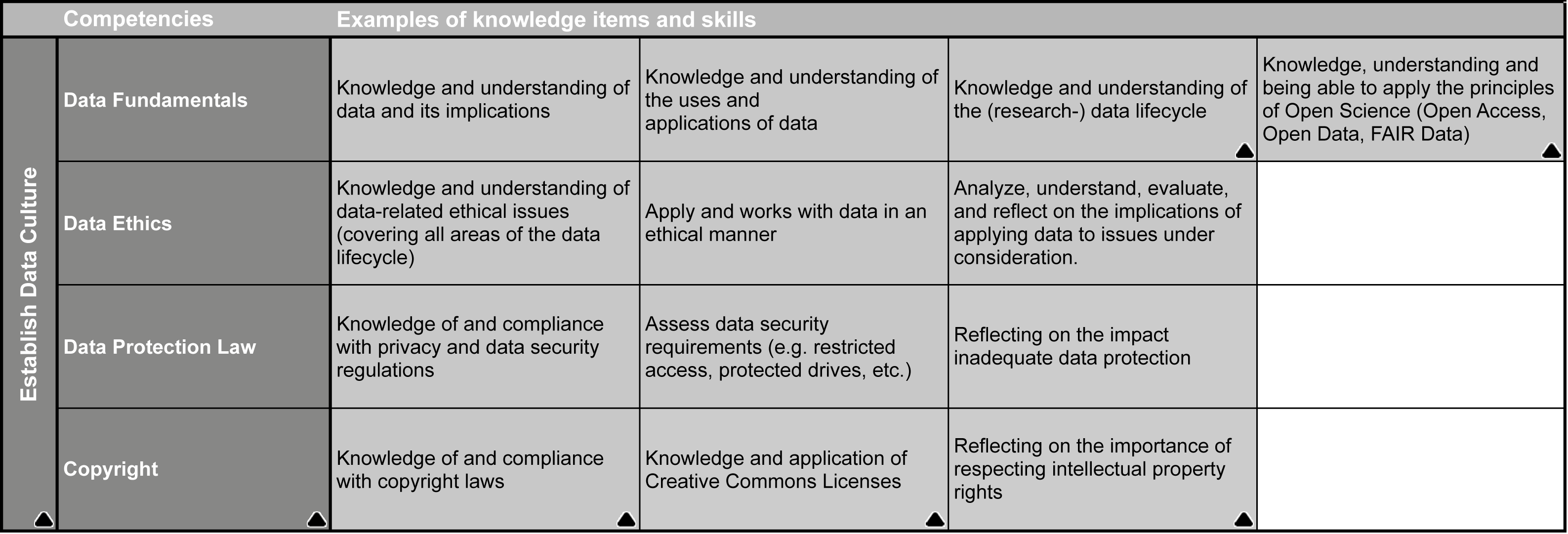}}
  \caption{The illustration depicts the 'Establish Data Culture' competence area, showcasing associated competencies, with examples of tasks/skills and knowledge items. The small triangles mark special features or innovations newly introduced with the proposed DaLI Competence Model.}
  \label{fig:establishDataCulture}
\end{figure}

Figure~\ref{fig:establishDataCulture} lists all competencies for this competence area with some examples of related knowledge items and tasks.

\subsection{Provide Data}\label{sec:Provide}
The 'Provide Data' competence area covers all the skills needed to provide meaningful data for a scientific study. This begins with determining what data is needed and ends with ensuring the quality of the data. 

Therefore, the “Provide Data” area of the DaLI Competence Model consist of: “Design Data Analysis”, “Identify Data Sources”, “Explore and Find Data”, “Verify Data Quality”, “Collect Data”, “Prepare data (formally)” and “Integrate data”. 

The competence “Design Data Analysis” encompasses the comprehensive knowledge and skills necessary to create a robust research design. It guides the transformation of a research question into an empirical investigation of statistically verifiable hypotheses. Key elements include defining the sample (or objects of study), selecting measurement instruments, and choosing appropriate research methods [siehe Literatur von Simone - Referenzen fehlen noch]. This area also encompasses knowledge of both quantitative and qualitative research methodologies~\cite{Roebken2022} and how to combine them effectively.
This competence precedes all other competencies in this area, as it provides the basic knowledge needed to collect the data meaningfully.

Building on the foundation laid by 'Design Data Analysis', the competencies 'Identify Data Sources', 'Explore and Find Data', 'Verify Data Quality', and 'Collect Data' focus on the next steps in the research process. Once the research requirements and data needs have been clearly defined through a robust research design, the next step is to identify trustworthy data sources, search for relevant datasets, or, if necessary, collect new data. The 'Identify Data Sources' competency focuses on selecting relevant and reliable sources of information. 'Explore and Find Data' covers locating existing datasets and determining their suitability for the research question. 'Verify Data Quality' assesses data's accuracy, completeness, and reliability to ensure it meets the standards necessary for effective analysis and valid research outcomes. 'Collect Data' addresses the process of obtaining new data, whether through surveys, experiments, or observations.

This competence area concludes with the competences 'Verify Data Quality', 'Prepare Data (formally)', and 'Integrate Data'. Once data is available, the next step is to assess its quality to ensure it meets the standards for reliable analysis. The 'Verify Data Quality' competence focuses on identifying any issues or inconsistencies in the data. 'Prepare Data (formally)' involves the formal processing and cleansing of the data, which can improve data quality by standardizing formats and correcting formal errors. Finally, 'Integrate Data' becomes essential when data from multiple sources needs to be combined into a single, cohesive dataset. This competence ensures that disparate datasets are accurately merged, providing a unified foundation for further analysis.

In many models, the competencies "Prepare Data (formally)" and "Integrate Data" are typically categorized under data management. In the DaLI Competence Model, however, these competencies are included in the "Provide Data" area due to the distinction between general formal preparation steps and specific preparation for data analysis.
General formal preparation, such as standardizing different data formats or correcting errors, is essential to make the data fundamentally usable, regardless of its subsequent application. Also, this step can be performed without knowledge of the subsequent processing steps. This split approach allows for flexible data preparation that is not tied to a specific analysis, ensuring consistency and reliability when integrating different datasets.
Data integration is also independent of the intended analysis. It can be carried out to supplement the data, possibly in relation to a specific question, thereby increasing the relevance and the value of the dataset in addressing the research questions.

\begin{figure}[!htb]
  \centerline{\includegraphics[width=\textwidth]{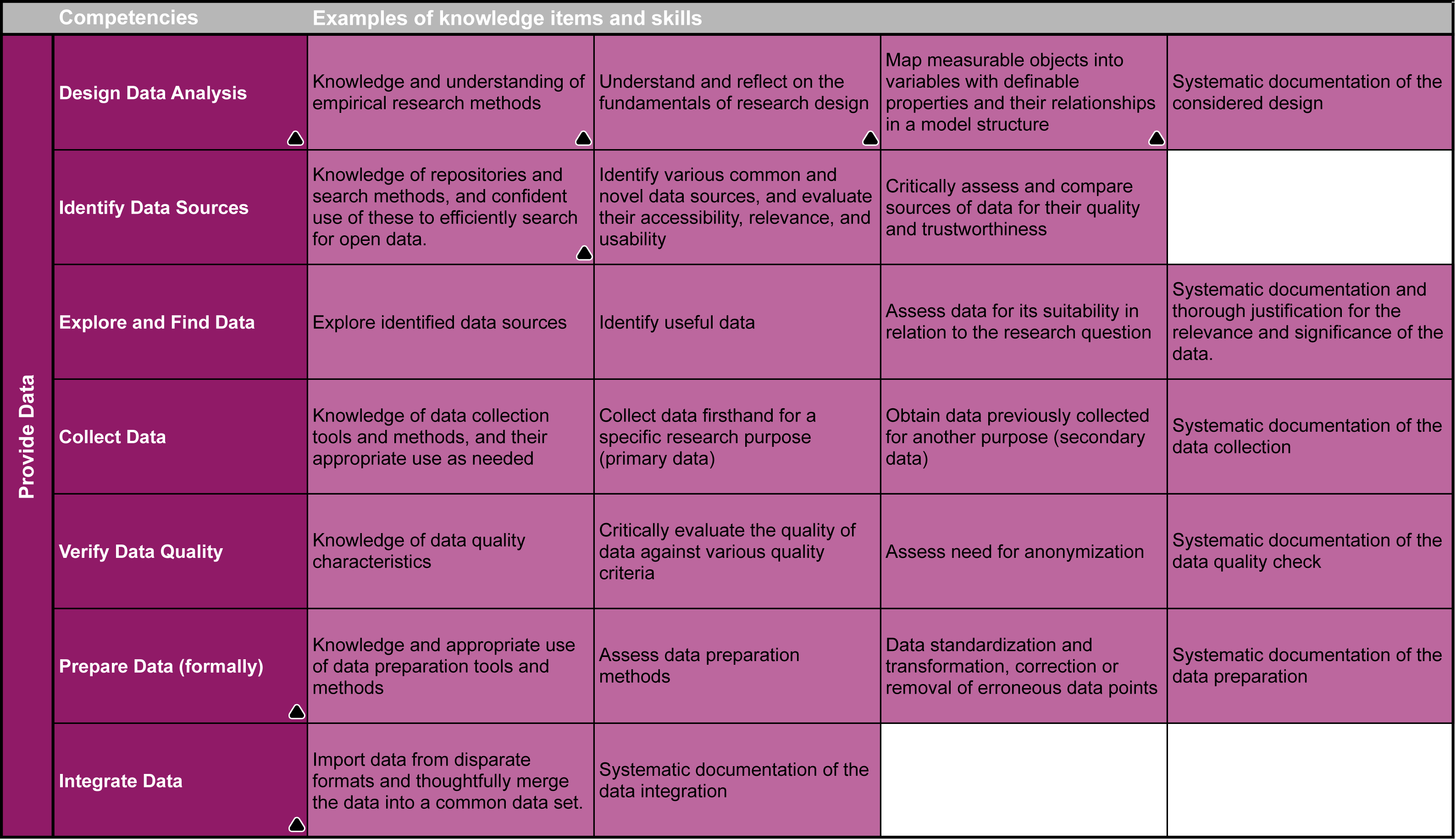}}
  \caption{The illustration depicts the 'Provide Data' competence area, showcasing associated competencies, with examples of tasks/skills and knowledge items. The small triangles mark special features or innovations newly introduced with the proposed DaLI Competence Model.}
  \label{fig:collectData}
\end{figure}

The early completion of these steps provides a solid foundation for the subsequent content-specific data preparation, which is addressed in the “Evaluate Data” competence area and tailored to the specific needs of the analysis.
Figure~\ref{fig:collectData} lists all competencies for this competence area with some examples of related knowledge items and tasks.

 \newpage
\subsection{ManageData}\label{sec:manageData}
The 'Manage Data' competence area encompasses all essential elements to ensure data remains consistently usable within the project and beyond. This competence area, therefore, takes into account the competencies: 'Organize Data', 'Create and Use Metadata', and 'Convert Data'. 

The competencies 'Organize Data' and 'Create and Use Metadata' ensure that data is stored in a structured way and accompanied by metadata to guarantee their findability, usability, and interpretability. Structuring makes data organized and easy to find, while metadata provides important contextual information, such as the origin of the data or any modifications made to the data over time.
'Convert Data' enables data conversion into various formats, i.e., to allow for data processing in different programs or transitioning between proprietary and open formats.

The competencies presented in this competence area were also covered by previous models, categorized under data organization and data management. The proposed competence model merely adds systematical documentation of the data organization and data conversion to ensure consistent traceability, reproducibility, and long-term usability of the work and data. A systematically documentation of the data structure and the processing steps facilitates collaboration with others, simplifies data sharing, enhances efficiency, and improves the quality and transparency of research and data analysis.

Overall, proper Data Management enables efficient searching and utilization of the data and the use of the data over a longer period of time. It lays the foundation for future use cases and sustainable reuse of the data.

\begin{figure}[!htb]
  \centerline{\includegraphics[width=\textwidth]{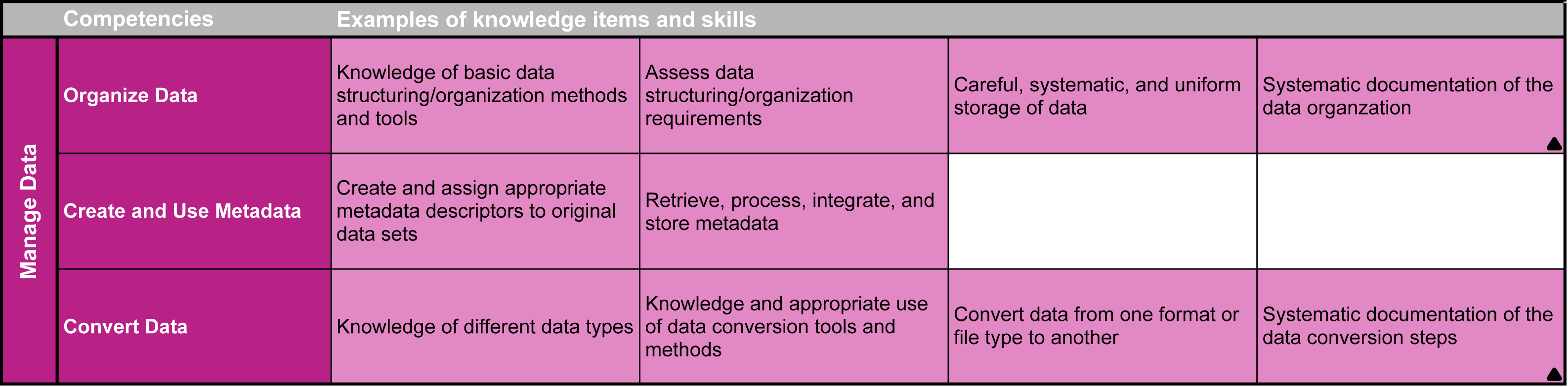}}
  \caption{The illustration depicts the 'Manage Data' competence area, showcasing associated competencies, with examples of tasks/skills and knowledge items. The small triangle marks the documentation related tasks that have been newly introduced with the proposed DaLI Competence Model.}
  \label{fig:manageData}
\end{figure}

Figure~\ref{fig:manageData} lists all competencies for this competence area with some examples of related knowledge items and tasks.

 \newpage
\subsection{Analyze Data}\label{sec:evaluateData}

The competence area "Analyze Data" encompasses all the competencies required to evaluate existing data in a statistical analysis or scientific study. While data analysis is a key component, proficiency in various analytical methods, familiarity with relevant software tools, and the ability to carefully prepare the data for the intended processing are also essential for effective evaluation.

In the competence area "Evaluate Data", the DaLI Competence Model, therefore, considers the topics "Tools and Programming Languages", "Process Data", "Quantitative Analysis", and "Qualitative Analysis" to provide a robust and comprehensive approach to evaluation. These topics ensure that evaluation is strongly emphasized and addressed holistically.

The competence "Tools and Programming Languages" involves the knowledge, selection, and application of appropriate data analysis tools and programming languages. This competence ensures that individuals can effectively utilize adequate software tools or programming languages to perform data analyses.

The competence "Process Data" focuses on preparing data for specific analyses. It involves steps that enhance evaluation outcomes, such as handling outliers and imputing missing data. This competence is essential to ensure the dataset is ready for detailed analysis.

The competencies "Qualitative Analysis" and "Quantitative Analysis" focus on the actual evaluation of data. Quantitative analysis utilizes analytical methods from fields such as statistics and machine learning, whereas qualitative analysis requires content-analytic methods~\cite{Mayring2014, Mayring2019, kuckartz2023qualitative}. Together, these competencies aim to provide a comprehensive overview of the diverse methods available and empower effective data evaluation.

Together, these competencies in the "Evaluate Data" area ensure that data is analyzed and interpreted effectively, allowing for the extraction of meaningful insights and conclusions. They support a structured approach to data evaluation, accommodating diverse research methodologies and analytical needs.

\begin{figure}[!htb]
  \centerline{\includegraphics[width=\textwidth]{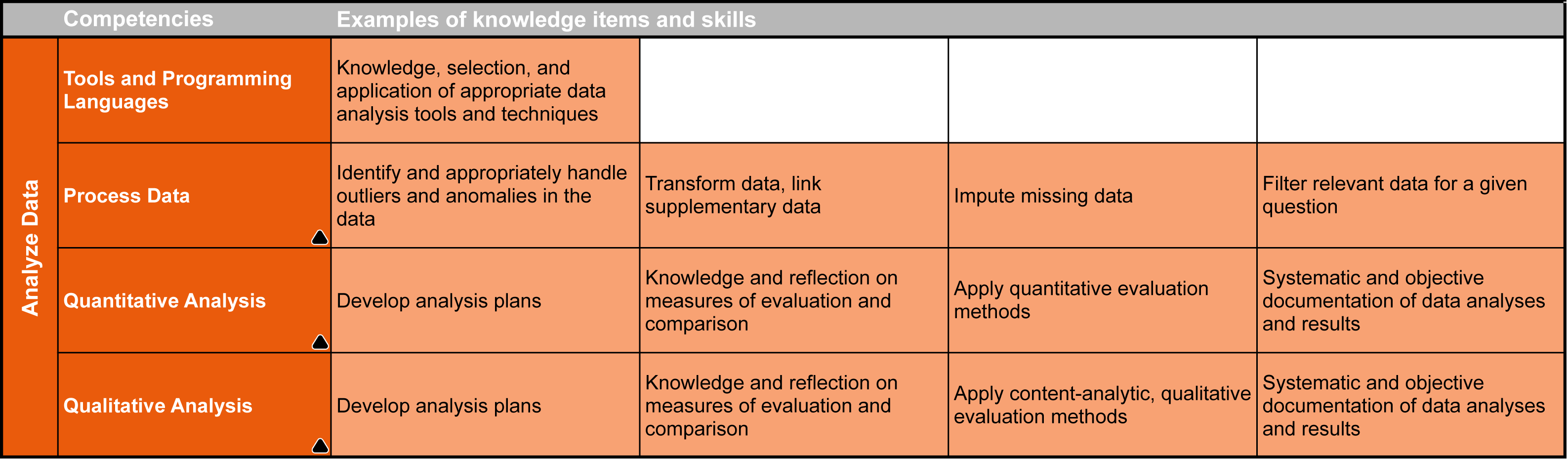}}
  \caption{The illustration depicts the 'Evaluate Data' competence area, showcasing associated competencies with examples of tasks/skills and knowledge items. The small triangles mark new features or innovations with the proposed DaLI Competence Model.}
  \label{fig:analyzeData}
\end{figure}

Competences such as "Data Visualization", "Verbal Data Presentation," and "Data-Driven Decision Making (DDDM)" are sometimes also found under Data Evaluation Areas~\cite{Ridsdale2015,UniGoettingen}. In the DaLI Competence Model, these competences are allocated to the areas of "Interpret Data"  and "Understand Data" in order to give them more space and to allow for more in-depth consideration.

Figure~\ref{fig:analyzeData} lists all competencies for this competence area with some examples of related knowledge items and tasks.

\subsection{Evaluate Data}\label{sec:interpretData}
The competence area "Evaluate Data" involves all the steps required to present and objectively evaluate the data and analysis results. Of course, this includes correct assessment of statistical analyses, visualization of data and analysis results, and presentation of any findings verbally. In addition, the ability to read and understand visual or verbal presentations given by others is crucial.
Therefore, the competence area "Evaluate Data" of the DaLI Competence Model consist of: "Visualize Data", "Understand Visualizations", "Evaluate Results", and "Present Data \& Results (Verbally)". 

The competence "Visualize Data" is a prerequisite for transforming data and analysis results into insightful visual narratives. This involves not only the knowledge, selection, and appropriate use of data visualization tools and methods but also an understanding of design principles and aesthetics to create informative and appealing visualizations. Additionally, this competence includes the critical evaluation of visualizations for accuracy, avoiding misrepresentation, and maintaining graphical integrity.

The competence of "Understand Visualizations" is crucial for extracting and understanding the insights embedded within graphical data presentations from arbitrary sources. A methodological understanding, similar to that required to create visualizations, is also needed to identify the core messages within these visualizations and to evaluate the choices made in their design critically.

The "Evaluate Results" competence focuses on the ability to evaluate data and related analytical results in relation to specific research questions and hypotheses. This includes identifying causal relationships, correlations, and statistical significance based on statistical analyses supported by appropriate visualizations. It requires a critical mindset to question and potentially falsify hypotheses, ensuring that conclusions drawn are supported by evidence. The competence of "Interpret Results" is crucial to derive meaningful insights and thus make informed decisions.

\begin{figure}[!htb]
  \centerline{\includegraphics[width=\textwidth]{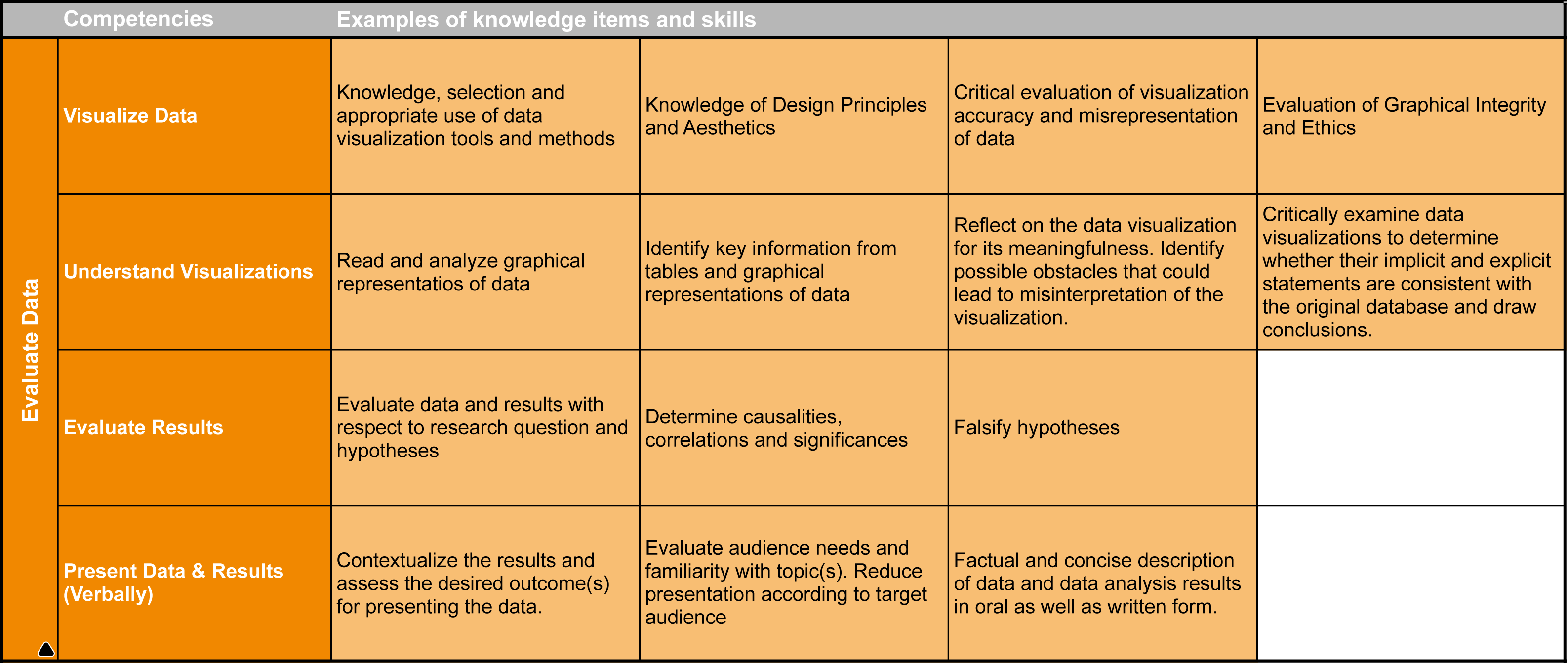}}
  \caption{The illustration depicts the 'Interpret Data' competence area, showcasing associated competencies with examples of tasks/skills and knowledge items. The small triangle indicates that this competence area has been newly introduced with the proposed competence model.}
  \label{fig:evaluateData}
\end{figure}

The competence "Present Data \& Results (Verbally)" involves effectively transforming analytical results into written or spoken words that communicate findings clearly to an audience. This requires the ability to distill complex information into straightforward and understandable language while maintaining the essence of the analysis. Understanding and accurately using appropriate terminology ensures clarity and precision in communication. Additionally, presentations should contextualize the results within the broader research objectives or real-world applications and must be tailored in detail and scope to suit the target audience.

The competences presented here, along with their associated skills and knowledge items, are also included in previous models. The main innovation is the consolidation of these competences under the new competence area "Interpret Data."
This restructuring allows for specialization and focus, and facilitates a more organized approach to teaching and learning by distinctly separating the interpretation of data from evaluation and analysis. 

Figure~\ref{fig:evaluateData} lists all competencies for this competence area with some examples of related knowledge items and tasks.
\subsection{Interpret Data}\label{sec:understandData}
The "Interpret Data" competence area offers a broad perspective that goes beyond the precise assessment, visualization, and presentation of data, as emphasized in "Evaluate Data." It focuses on the critical evaluation and contextual understanding of data to recognize its limitations, ensuring both data integrity and methodological accuracy. This approach facilitates a deeper comprehension of results, enabling the derivation of actionable conclusions that are statistically sound and contextually informed. 
Therefore, the competence area "Interpret Data" consist of: "Examine and Apply Data", "Contextualize Data", and "Derive Actionable Information".

The competence "Examine and Apply Data" refers to the process of critically assessing data and the results of analyses to ensure they are reliable and valid. It includes a broad awareness of high-level issues and challenges associated with data, such as potential limitations or biases. This competence involves recognizing these limitations in data, results, and conclusions. 

"Contextualize Data" involves situating analysis results within the broader framework of existing research and disciplinary knowledge. This process allows scientists to compare their findings with related studies, to validate their results, and to understand their contribution to the field. 

The competence "Derive Actionable Information" focuses on translating the above insights and understandings into practical outcomes. This encompasses the ability to derive conclusions and identify possibilities for actions based on data insights. Evaluating the effect of potential future actions is part of this competence, as is recommending strategic actions grounded in data analysis results.

The competences presented here, along with their associated skills and knowledge items, are also included in previous models. The main innovation is the consolidation of these competences under the new competence area "Interpret Data". 

\begin{figure}[!htb]
  \centerline{\includegraphics[width=\textwidth]{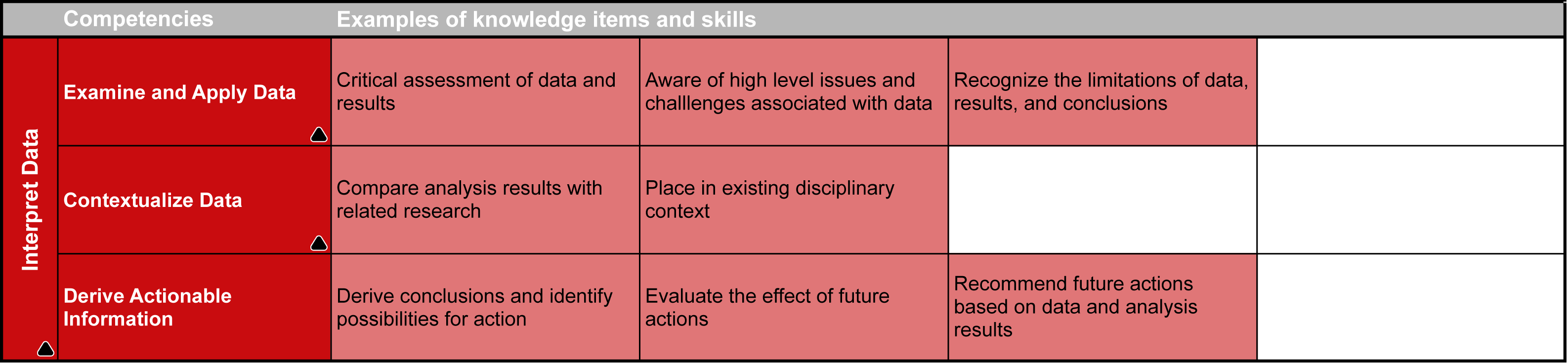}}
  \caption{The illustration depicts the 'Understand Data' competence area, showcasing associated competencies with examples of tasks/skills and knowledge items. The small triangles mark new features or innovations, introduced with the proposed DaLI Competence Model.}
  \label{fig:interpretData}
\end{figure}

\subsection{Publish Data}\label{sec:publishData}

The last competence area, "Publish Data", completes the project lifecycle. Publication of data and research findings typically signifies the conclusion of a research project. However, finalizing a data project entails more than mere publication; annotated data must be archived and preserved for future use, secondary data must be appropriately cited, and licensing arrangements must be carefully determined. Additionally, knowledge of appropriate publication opportunities is needed for the publication of reputable scientific work. Therefore, the competence area "Publish Data" consists of "Store and Maintain Data", "Preserve Data for Future Reuse", "Cite (Research) Data Sets" and "Share Data (Open Data)".

The competence "Store and Maintain Data" is the ability to systematically store, manage, and maintain research data throughout its lifecycle to ensure accessibility, integrity, and security. This competence involves selecting appropriate storage solutions, implementing best practices for data preservation, and ensuring compliance with institutional, legal, and ethical requirements. It also includes strategies for long-term data maintenance, backup, and controlled access to facilitate collaboration from multiple locations and future reuse.

The competence "Preserve Data for Future Use" focuses on making data reusable beyond the original research context by applying structured preservation methods (cf. FAIR Principles~\cite{Wilkinson2016,NG2017a}), metadata standards, and sustainable storage solutions. It involves systematically preparing, documenting, and storing research data to ensure its usability, accessibility, and integrity in the long term. Additionally, it requires considering future technological developments, interoperability, and the requirements of repositories and archives to maximize the longevity and impact of research data.

The "Cite (Research) Data Sets" competence ensures that secondary research data is cited properly, preserving data sources' academic value and traceability.

The competence "Share Data (Open Data)" is the ability to share research data and results while ensuring accessibility, reproducibility, and compliance with ethical, legal, and institutional guidelines. This competence involves selecting appropriate data-sharing platforms, applying open data licensing frameworks, and using persistent identifiers such as DOIs to enhance discoverability and citation. 

\begin{figure}[!htb]
  \centerline{\includegraphics[width=\textwidth]{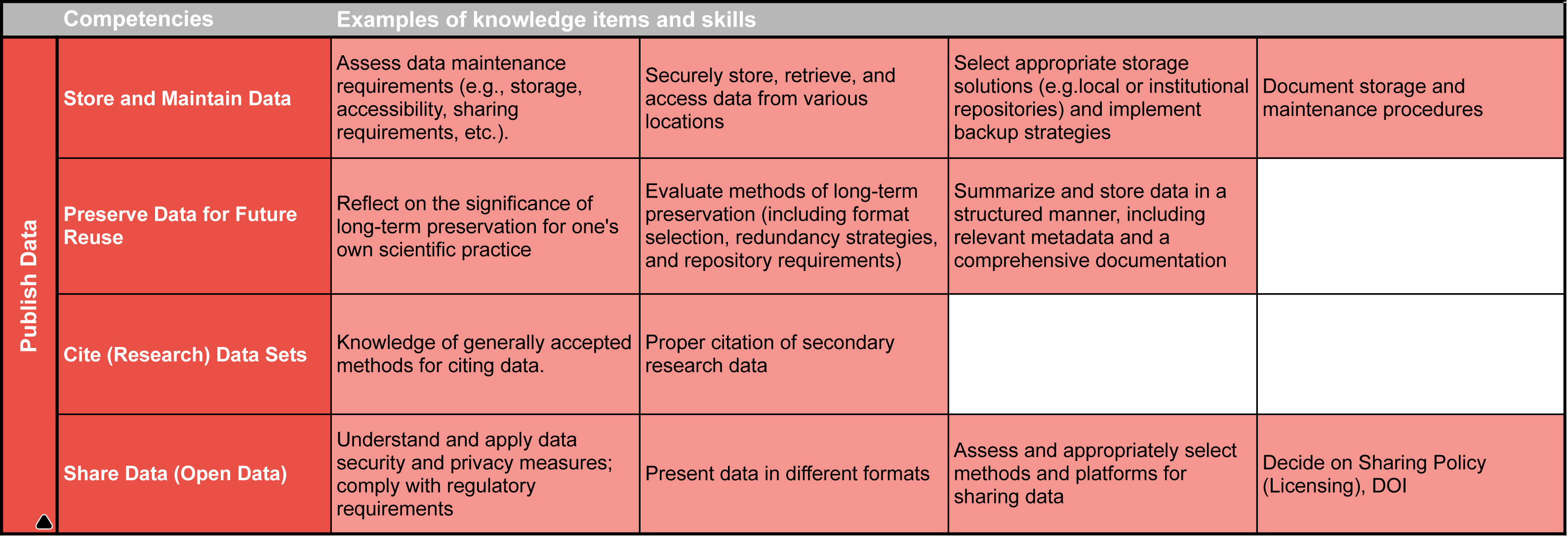}}
  \caption{The illustration depicts the 'Publish Data' competence area, showcasing associated competencies with examples of tasks/skills and knowledge items. The small triangle indicates that this competence area has been newly introduced with the proposed competence model.}
  \label{fig:publishData}
\end{figure}

The competences presented here, along with their associated skills and knowledge items, are also included in previous models. The key innovation lies in the consolidation of these competences under the new competence area, "Publish Data." As emphasized earlier, appropriate archiving, publication, and the resulting chance of reusing data are especially important in research and higher education, as they ensure transparency, reproducibility, and long-term accessibility of knowledge.
This restructuring enables specialization and focus while emphasizing the importance of open science and reproducibility. Figure~\ref{fig:publishData} lists all competencies for this competence area with some examples of related knowledge items and tasks.

\section{Conclusion and Future Work}\label{sec:conclusion}

As outlined in the introduction, the existing definitions of data literacy vary widely, making it challenging to fully understand and effectively communicate the concept of data literacy. This paper aims to develop a clear and comprehensive definition of 'data literacy' and establish a well-defined competence model that allows for conveying and recognizing data literacy skills within educational settings and certification systems. The envisioned model provides a structured approach to teaching data literacy, ensuring that it is ethically and legally sound while addressing the specific needs of higher education and research.

The proposed DaLI Competence Model is a structured framework designed to foster data literacy in higher education and research. The model is organized around seven key competence areas: 'Establish Data Culture', 'Provide Data', 'Manage Data', 'Analyze Data', 'Evaluate Data', 'Interpret Data', and 'Publish Data'. These areas align with the research data lifecycle, ensuring the competencies acquired are theoretically sound and practically applicable in academic and professional settings.

The DaLI Competence Model strongly emphasizes establishing a responsible data culture, integrating ethical, legal, and social considerations into every stage of data handling. This includes a dedicated focus on data ethics, data protection law, and intellectual property rights, ensuring that users have the ethical insights, legal knowledge, and technical capabilities to work with data responsibly and understand the broader impact of their actions.

The detailed breakdown of competencies into seven distinct areas allows for a stronger focus on these domains. Thus, the model provides a clear structure for designing courses, assessing competencies, and recognizing achievements within general educational settings and certification systems. By grouping the relevant competencies into the new areas of "Evaluate Data", "Interpret Data", and "Publish data", the model especially emphasizes understanding, critical questioning, and open sharing of data, results, and insights. In doing so, the DaLI Competence Model promotes a responsible and open approach, which is particularly required in higher education and research.

Future work will involve implementing and evaluating the DaLI Competence Model in various educational settings. This will include developing specific course modules, training materials, and assessment tools tailored to the different competence areas. 
Additionally, future work will focus on adapting and enhancing the DaLI Competence Model to address emerging topics in artificial intelligence (AI). As AI technologies evolve and become increasingly integrated into data practices, we see the need to consider the required skills to deal with data responsibly in the context of AI. This includes understanding AI algorithms, addressing issues of bias and transparency, and fostering skills in responsible AI development and deployment. By ensuring that these elements are effectively woven into the framework, the DaLI Competence Model will remain relevant and robust, equipping students, lecturers, teachers, and researchers with the tools they need to navigate the complex landscape of data and AI.

\section*{Acknowledgments}

We thank Prof. Phillip Heidkamp for his fruitful ideas and the consistent, constructive collaboration. 
This work was also financially supported by the \textit{Stifterverband der deutschen Wirtschaft e.V.} as part of the \textit{Data Literacy Education.NRW} program, and by TH Köln, whose continuation of the initial funding helped to establish the DaLi-Initiative as a lasting institution for promoting data literacy across the university.

\bibliographystyle{abbrv} 
\bibliography{bibliography}  

\end{document}